\pdfoutput=1

\documentclass[preprint]{aastex6}

\newcommand{\be}{\begin{equation}}
\newcommand{\ee}{\end{equation}}
\newcommand{\diff}{\ \mathrm{d}}
\newcommand{\sdiff}{\mathrm{d}}
\newcommand{\del}{\partial}

\newcommand*{\thead}[1]{\multicolumn{1}{c}{#1}}
\newcommand*{\tunit}[1]{\multicolumn{1}{c}{$(\SI{}{#1})$}}

\shorttitle{A spatially extended SSC model}

\usepackage{savesym}
\savesymbol{tablenum}
\usepackage[exponent-product=\cdot]{siunitx}

\usepackage{amsmath}

\begin{document}


\title{A numerical model of parsec scale SSC morphologies and their radio emission}


\author{S. Richter\altaffilmark{1} and F. Spanier}
\affil{Centre for Space Research\\
North-West University\\
2520 Potchefstroom, South Africa}


\altaffiltext{1}{Stephan.Richter@nwu.ac.za}

\begin{abstract}
In current models for jets of AGNs and their emission a shortcoming in the description and understanding of the connection between the largest and smallest scales exists. In this work we present a spatially resolved SSC model extended to parsec scales, which opens the possibility of probing the connections between the radio and high energy properties. We simulate an environment that leads to Fermi-I acceleration of leptonic particles and includes the full time dependence of this process. Omitting the restriction of a finite downstream region, we find that the spectral energy distribution (SED) produced by the accelerated particles strongly depends on their radial confinement behind the shock. The requirement, for both the restriction of high energy emission to a small region around the shock and the production of a flat radio spectrum, is an initial linear increase of the radius immediately behind the shock, which then slows down with increasing distance from the shock. A good representation of the data for the Blazar \textit{Mkn501} is achieved by a parameterized log-function. The prediction for the shape of the radio blob is given by the flux distribution with respect to shock distance.

\end{abstract}

\keywords{acceleration of particles -- BL Lacertae objects: individual (Mrk501) -- galaxies: jets -- radiation mechanisms: non-thermal -- relativistic processes}



\section{Introduction} \label{sec:introduction}

Radio-loud active galactic nuclei (AGN) are one of the showpieces of the multi-messenger approach. Their spectral energy distribution (SED) covers a large range of observatories from radio antennas to Cherenkov telescopes. Although qualitatively they have a common SED shape, their individual spectra span a wide parameter space \citep{2011ApJ...740...98M}. Nevertheless a large database of spectra and lightcurves in various bands as well as correlations between them~\citep[e.g.][]{2008Natur.452..966M,2011A&A...532A.146L} can be used to constrain models of acceleration and radiation processes.
In the case of Blazars, at least for the so called high peaked BL Lacs (HBLs), the synchrotron self Compton models \citep[SSC, ][]{1991ApJ...377..403C} are very successful in describing almost the entire SED with a small number of parameters \citep[e.g.][]{2011ApJ...727..129A,2011ApJ...736..131A}. For low peaked sources and flat spectrum radio quasars (FSRQs), a hybrid model that includes synchrotron radiation and photohadronic processes due to the presence of non-thermal protons \citep[e.g.][]{2013EPJWC..6105009W} or the consideration of external radiation \citep[][]{2002ApJ...564...86B} might produce better representations of the data. Although the observation of ultra short variability \citep[e.g][]{2007ApJ...664L..71A,2011ICRC....8..171P} challenges the self-consistent picture, the radiation mechanism is quite robust as long as the size of the emission region can be set sufficiently small.

The difficulty then arises from the explanation of acceleration timescales and boundary conditions. The discrepancy between the observed variability timescale and the light crossing time of the black hole ergosphere is an additional one, which is not discussed here. The problem discussed here is that of the limited size of the emission region, which makes it impossible to explain the radio observations. Those observations, even for very large baseline interferometry (VLBI), take place at much larger scales~\citep[e.g.][]{2001ARA&A..39..457K}. Models of the jet morphology on larger scales are based on (general relativistic) magnetohydrodynamic (MHD) methods, but those models can only produce synthetic spectra~\citep{2009ApJ...695..503G,2011ApJ...737...42P}.  Furthermore they are scale invariant and can not explain the necessary and observed length scales, respectively. Existing spatially resolved models ~\citep[e.g.][]{2011MNRAS.416.2368C,2008ApJ...689...68G} are focusing on variability patterns and do not extend to the length scale of radio blobs.

Despite these limitations, radio observations yield the highest resolution of the jet morphology and might be used to infer the origin of the very high energy (VHE) radiation, if a significant time correlation to VHE flares can be observed. In order to exploit the radio band and its correlations, first a physically motivated connection between the two scales has to be established. Any such approach should be able to explain both the radio properties at the parsec scale (i.e. morphology and spectral index) and the confinement of the VHE emitting particles to a sufficiently small region. Such a model could then be constrained by time correlations between various bands.

In section~\ref{sec:model} we present the details of an extension to the often used homogeneous SSC model. In our approach, we connect the acceleration with a representation of a shock in one spatial dimension. This dimension can then be extended up to parsec scales. In section~\ref{sec:results} we summarize the effect of various jet-morphologies on the overall SED. The fits are, in addition to the high energy data, constrained by the spectral index of the radio part of the SED. One obtains the spatial flux distribution, which can be connected to VLBI radio maps. Section \ref{sec:discussion} will summarize our results. To add to this, numerical and conceptional limits of the current model as well as future work will be discussed.

\section{Model}
\label{sec:model}
The characteristic double hump structure found in the SEDs of AGN can be explained elegantly with the SSC paradigm. Starting from a (broken) powerlaw distribution for the leptonic particle content, fits for at least high peaked BL Lacs (HBLs) are generally possible. The additional lightcurves available for many sources favor a time dependent approach, where the used electron distribution is not set a priori, but emerges from the same set of parameters used for the computation of radiation processes \citep[e.g.][]{2010ASTRA...6....1W}. Since the efficiency of the acceleration influences both the spectral index of the electron (and hence of the photon) distribution and the rise time of flares it is possible to further constrain SED fits from lightcurve data. Furthermore it is possible to learn something about the nature of the acceleration process.

The process usually assumed for the fast production of a powerlaw distribution is the so called Fermi-I acceleration ~\citep[e.g.][]{2004PASA...21....1P}. This process is based on elastic particle scattering in the vicinity of a MHD shock. Acceleration arises then during the isotropization of the particle distribution after the shock crossing. This process can be seen as diffusion in the rest frame of the ambient plasma, which itself is streaming away from the shock. Consequently the process is non local and can be incorporated into a spatially resolved model.

For easier reference all quantities used in this work are summarized in Table~\ref{tab:quantities}.

\begin{table}[h]
	\centering
	\caption{Summary of used quantities.}
	\begin{tabular}{cl}
		\tableline\tableline
		\thead{quantity} & \thead{description}\\
		\tableline
		\\[-5mm]
		$z$ & spatial coordinate along the jet axis and parallel to the shock normal\\
		$\mu$ & $\mu=\cos\theta$ with the pitch angle $\theta$\\
		$\gamma$ & particle Lorentz factor\\
		$\nu$ & photon frequency\\
		$R$ & radius, dependent on $z$, of the simulated volume\\
		$V_S$ & shock speed\\
		$B$ & magnetic field strength\\
		$r$ & shock compression ratio\\
		$D$ & momentum diffusion coefficient\\
		$\mathcal{D}$ & Doppler factor\\
		$\mathcal{Z}$ & redshift\\
		\tableline
	\end{tabular}
	\label{tab:quantities}
\end{table}

\subsection{Geometry}
The simulation box is discretized in the $z$ direction along the shock normal and assumed homogeneous in the perpendicular plane. This allows the movement of particles across the shock, while keeping the computational cost at a minimum. We restrict ourselves to mildly-relativistic and non-oblique shocks. The simulated system can therefore be thought of as a non-relativistic shock within a plasma blob, that is moving along the jet with a relativistic Doppler factor. A generalization towards a time-dependent description of relativistic shocks would be desirable, but is not in the realm of current numerical possibilities. Also the authors are not aware of such a model or implementation in the literature. However, steady state descriptions of relativistic shocks exist~\citep[e.g.][]{1999JPhG...25R.163K} and yield much steeper spectra than usually observed.

\begin{figure}[ht]
  \centering
  \includegraphics[width=0.5\textwidth]{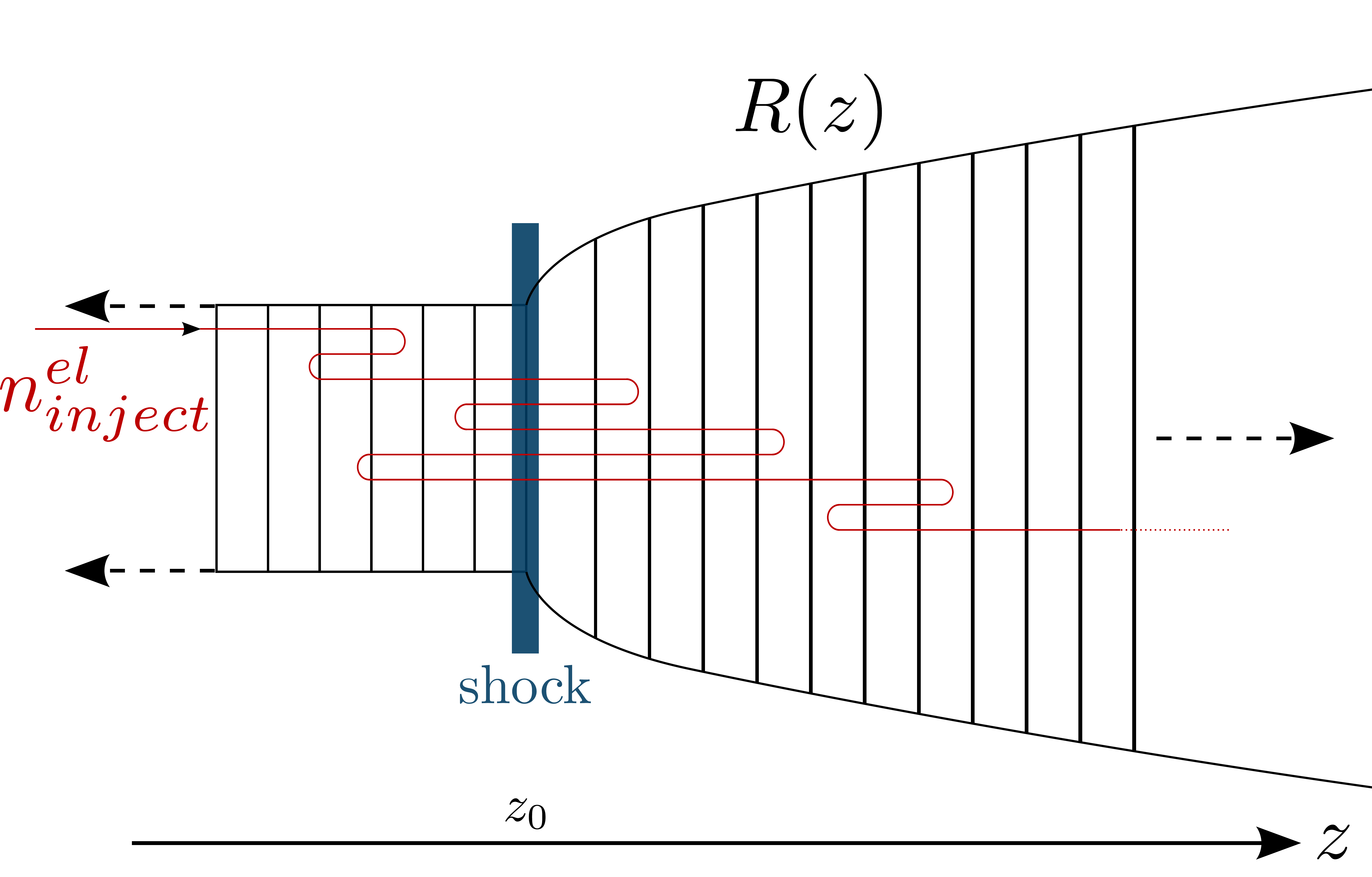}
  \caption{Schematic illustration of the used geometry. The spatial discretization is along $z$, the direction of the magnetic field and parallel to the shock normal. The shock is positioned explicitly at $z=z_0$. The upstream region is set homogeneous, while in the downstream region the bulk flow is expanding with an arbitrary function $R(z)$. Particles can scatter between the two distinct directions, but no explicit movement perpendicular to $z$ is computed. The red line is only intended to illustrate the scattering process.}
  \label{fig:geometrie}
\end{figure}

The setup is described schematically in Fig.~\ref{fig:geometrie} and is designed to represent the space at the acceleration zone and downstream of it with the goal of explaining the radio core and its properties, as proposed by~\citet{2008ASPC..386..437M}. Vertical lines divide the simulation box into cylindrical cells of variable radius $R$, which is a smooth function of $z$.

Every cell has its particular particle and photon content and, in principle, parameter set. A change in parameters along $z$ should be physically motivated though.
\subsubsection{Radial confinement}
\label{sec:radial_confinement}
As will be shown in section~\ref{sec:results}, homogeneous SSC models are generally introducing an artificial boundary condition that will alter the source emission below the synchrotron cooling break, especially in the radio regime. In order to replace the numerical boundaries with a physical boundary condition, we extend the spatial size $z_{max}$ of our simulation. The goal is to find a relation between the radius $R$ and the distance from the shock $z-z_0$ that produces the observed spectral index in the radio regime, as well as a radio emission that extends to VLBI scales.

A first ansatz for the relation $R(z)$ was tested in~\citep[][in the following \textit{RS}]{2013EPJWC..6105010R} in the form of a conical expansion. It was found that an expansion behind the shock and an opening angle of the order $\sim1$ or larger can explain the spatial confinement of the high energy emission. However, the results of \textit{RS} showed that a linear increase of $R$ will create neither the correct spectral index in the measured SEDs, nor radio emission in the form of a blob on the observed length scales.

Therefore, the initial expansion behind the shock has to slow down at larger distances, and should show only marginal increase at very large distances, which is consistent with observations by~\citet{1999Natur.401..891J}.

Both power-laws with variable exponents and log-functions were used as test functions to model this qualitative form. It was found that the best description of the data could be achieved with a log-function, whose first derivative at the shock position $z_0$ is set to $\alpha$:
\be
\label{eq:rofz}
  R(z)=\begin{cases} 	R_0 &\mbox{for } z\leqq z_0 \\
			R_0\left(1+\alpha\log\left(1+\frac{z-z_0}{R_0}\right)\right) &\mbox{for } z>z_0 \\\end{cases}
\ee
Although the model is not restricted to this function, it is used throughout the work presented here.
\subsubsection{Shock-front and particle advection}
The shock-front can be modeled by a jump of the streaming velocity of the ambient plasma. Given the properties (velocity $V_S$ and compression ratio $r$) of the shock the bulk velocities in the up- ($V_P^u$) and downstream ($V_P^d$) - as seen in the shock-frame - can be calculated via
\be
  V_P^u=-V_S\quad\text{and}\quad V_P^d=-\frac{V_S}{r}\quad.
\ee
Velocities throughout this work are denoted in units of the speed of light $c$. In the case of a single shock one can simply assign these two values to the bulk velocities in the grid cells in the shock-frame:
\be
\widetilde V_P(z)=\begin{cases}
			V_P^u & \text{if } z < z_0 \\
			V_P^d & \text{if } z > z_0
		  \end{cases}
\ee
All other calculations then take place in this frame of reference.

The scattering driving the acceleration process is assumed to be pitch angle scattering. Consequently the description of the particle distribution can not be isotropic. A full discretization of the pitch angle $\mu=\cos(\theta)$ is, however, numerically expensive. Since the deviation from isotropy will only be very small, we follow a less sophisticated approach. The particle distribution $n(z,\gamma,\mu)$ is divided into two half-spheres. Their boundary is parallel to the shock plane and in the plasma rest frame one can define:
\be
  n^+(z,\gamma)=\int_{0}^{1}\!n(z,\gamma,\mu)\diff\mu\ ,\qquad n^-(z,\gamma)=\int_{-1}^{0}\!n(z,\gamma,\mu)\diff\mu
\ee
The ratio between $n^+$ and $n^-$ is a measure for the anisotropy of the distribution. The pitch angle scattering is parametrized by a fraction $W$ of the particle density that is scattered from one half sphere into the other. In the shock frame, $W$ has to be calculated per direction since there will be a net flux towards the downstream. The values of $W^\pm$ are chosen as such that the isotropic case is the equilibrium between $n^+$ and $n^-$:
\be
  \frac{W^-}{W^+}=\frac{n_{iso}^+(\widetilde V_P)}{n_{iso}^-(\widetilde V_P)}=\frac{1+\widetilde V_P}{1-\widetilde V_P}\quad
\ee
This expression can be obtained by taking an isotropic (in the plasma frame) and monoenergetic distribution in momentum space and boost it into the shock frame. The corrected momentum space volume can be obtained in spherical coordinates
\be
  n^+\propto\int_{-\widetilde V_P}^1 \sdiff \mu\int_0^{2\pi} \sdiff\phi=2\pi(1+\widetilde V_P),\ n^-\propto\int_{-1}^{-\widetilde V_P} \sdiff \mu\int_0^{2\pi} \sdiff\phi=2\pi(1-\widetilde V_P)\quad,
\ee
using the approximation $V_{part}=1$ (since $\gamma\gg1$) for the speed of the particles. The lower boundary of the integral represents particles whose momentum along the $z$-direction vanishes after the boost into the shock frame. The average scattering rate $\sqrt{W^+W^-}=W=t_{iso}^{-1}$ can be seen as an inverse isotropization timescale and enters the model as a parameter.

The average advection speed of the particles in each cell is computed from the velocity of the bulk plasma in the shock frame $\widetilde V_P$, again using $V_{part}=1$:
\be
 \label{eq:vadvp}
  V_{adv}^+=\frac{1}{1+\widetilde V_P}\int_{-\widetilde V_P}^{1}\!\frac{\widetilde V_P+\mu}{1+\widetilde V_P\mu}\diff\mu
		=\frac{\widetilde V_P - (\widetilde V_P-1) \ln(1 - \widetilde V_P)}{\widetilde V_P^2}
\ee
The lower boundary of the integral is the angle for which the speed of the particle's gyro-center in the shock-frame $\mu+\widetilde V_P$ is zero.
Equivalently
\be
 \label{eq:vadvm}
  V_{adv}^-=\frac{1}{1-\widetilde V_P}\int^{-\widetilde V_P}_{-1}\!\frac{\widetilde V_P+\mu}{1+\widetilde V_P\mu}\diff\mu
		=\frac{\widetilde V_P - (\widetilde V_P+1) \ln(1 + \widetilde V_P)}{\widetilde V_P^2}
\ee
can be computed. These velocities then enter Eq.~\ref{eq:kinetic_el} and describe the advection of the particle distribution.

\subsection{Kinetic Equations}
The complete time evolution of the distribution of non-thermal particles is described by the kinetic equation~\ref{eq:kinetic_el}. It is derived from the Fokker-Planck equation and solved for every particle species and direction of flight. After the same integrations over $\mu$, one obtains:
\begin{multline}
  \label{eq:kinetic_el}
  \frac{\del n^\pm(z,\gamma)}{\del t}+\frac{\del}{\del z}\left(V_{adv}^\pm n^\pm(z,\gamma)\right)
	=W^\mp n^\mp(z,\gamma')-W^\pm n^\pm(z,\gamma)\\
+\frac{\del}{\del \gamma}\Biggl[D\gamma^2\frac{\del n^\pm(z,\gamma)}{\del\gamma}+\Bigl(P_{sync}(\gamma)+P_{IC}(\gamma)+P_{ad}(\gamma)-2D\gamma\Bigr)\ n^\pm(z,\gamma)\Biggr]+S(z,\gamma,t)
\end{multline}
The following process are included:

An advection term using the values $V_{adv}^\pm$ of eqs.~\ref{eq:vadvp},\ref{eq:vadvm}.

The two terms in the first line describe the scattering between $n^+$ and $n^-$, including an energy-change $\gamma=(\Gamma(\widetilde V_P))^2\gamma'$. Here $\Gamma(\widetilde V_P)$ is the Doppler factor for the boost into the plasma rest frame, in which the scattering is elastic.

The momentum diffusion represents the Fermi-II acceleration. The momentum diffusion coefficient $D=({v_A}^2)/(9\kappa_\parallel)$ depends on the Alfv\' en speed $v_A$ and the parallel diffusion coefficient $\kappa_\parallel$~\citep{1983ApJ...270..319W}. The linear term $-2D\gamma$ results from the transition to the isotropic description.

The synchrotron losses are calculated following~\citet{1965ARA&A...3..297G} as
\be
	\label{eq:synclosses}
	P_{sync}=\frac{1}{6\pi}\frac{\sigma_TB^2}{mc}\gamma^2=\beta_s\gamma^2\quad,
\ee
where $\sigma_T$ is the Thomson cross section.

The inverse Compton process, which depends on the photon distribution $N(\nu)$, and its cooling of the electron population are calculated with the full Klein-Nishina cross section $\sdiff N_{\gamma,\nu}/(\sdiff\nu'\diff t)$ \citep{1970RvMP...42..237B}:
\be
  \label{eq:pic}
  P_{IC}=\frac{1}{mc^2}\int\sdiff\nu'\ h\nu'\int\sdiff\nu\ N(\nu)\frac{\sdiff N_{\gamma,\nu}}{\sdiff\nu'\diff t}\quad.
\ee

The injection of particles at the edges of the simulation box are denoted by the function S. Since the acceleration of particles is computed self-consistently (i.e. within a physical environment that is also the source of the produced radiation) it is sufficient to assume only delta like (in energy) injections at the upstream edge. Throughout this work we use the constant injection
\be
  S(z,\gamma,t)=N_{inj}\ \delta(z)\delta(\gamma-\gamma_{inj})\quad.
\ee

\subsection{Adiabatic expansion}
\label{sec:adex}
The expansion behind the shock requires the adjustment of the cooling and the values of the magnetic field. The dependencies on the function $R(z)$ can be expressed by
\be
	P_{ad}=\frac{1}{3}\frac{\dot V}{V}\gamma=\frac{2}{3}\frac{\widetilde V_P}{R(z)}\frac{\del R}{\del z}\gamma\ ,\qquad B(z)=B_0\left(\frac{R_0}{R(z)}\right)^m\quad,
\ee
where $V$ and $\dot V$ are the volume of the expanding particle distribution and its time dependence, respectively. Since we assume a parallel shock and a flow along the direction of the magnetic field we choose $m=2$~\citep{2011ApJ...740...64L}. The adiabatic cooling power $P_{ad}$ enters Eq.~\ref{eq:kinetic_ph} as an additional term. Furthermore the dilution of the particle density has to be taken into account.

\subsection{Photon time evolution}
The back-reaction of the produced synchrotron photons via the inverse Compton process makes it necessary to simultaneously compute the photon distribution $N(z,\nu,t)$. It becomes time dependent and is integrated according to Eq.~\ref{eq:kinetic_ph}.
\be
  \label{eq:kinetic_ph}
  \frac{\del N(z,\nu)}{\del t}=-c\ \kappa_{\nu,SSA}\ N(z,\nu)+\frac{4\pi}{h\nu}(\epsilon_{\nu,IC}+\epsilon_{\nu,sync})-\frac{N(z,\nu)}{t_{esc}}\quad.
\ee
Since all radiation processes are calculated in the isotropic approximation, i.e. from $n(z,\gamma)=n^+(z,\gamma)+n^-(z,\gamma)$, and no external photon fields are considered, the gradient of the photon distribution is very small. Hence in this case photon propagation can be neglected. The synchrotron power spectrum
\be
  \epsilon_{\nu,sync}=\frac{1}{4\pi}\int\sdiff\gamma\ n(\gamma)\ P_\nu(\gamma,\nu)
\ee
is computed in the Melrose-approximation~\citep{1983Ap&SS..92..105B}:
\be
  \label{eq:melrose}
  P_\nu(\gamma,\nu)\approx1{,}8\frac{\sqrt3\ q^3B}{m\ c^2}\left(\frac{\nu}{\nu_c}\right)^\frac{1}{3} e^{-\frac{\nu}{\nu_c}}\quad.
\ee
Employing the same approximation the synchrotron self absorption coefficient
\be
 \kappa_\nu=\frac{1}{8\pi\nu^2m^2c^2}\int\sdiff \gamma\ \gamma^2 \frac{\partial}{\partial \gamma}\left(\frac{n(\gamma)}{\gamma^2}\right)P_\nu(\gamma,\nu)
\ee
is calculated \citep{1965ARA&A...3..297G}. As pointed out earlier, the inverse Compton process takes place in the Klein-Nishina regime. The resulting losses and gains $\epsilon_{\nu,IC}$ are obtained from Eq.~\ref{eq:icwins}
\be
  \label{eq:icwins}
  \epsilon_\nu=\frac{h\nu}{4\pi}\int\sdiff\gamma\ n(\gamma)\int\sdiff\nu'\left(\frac{\sdiff N_{\gamma,\nu'}}{\sdiff\nu\diff t}\ n(\nu')-\frac{\sdiff N_{\gamma,\nu}}{\sdiff\nu'\diff t}\ n(\nu)\right)\ .
\ee
The catastrophic loss is parameterized by the escape timescale $t_{esc}=R/c$ and depends on the radius of the cell.

The total flux emerging from all spatial cells is calculated similar to the model of~\citet{1979ApJ...232...34B}. The integrated flux per cell can be calculated via
\be
 \nu F(\nu,z)=\frac{\mathcal{D}^4}{1+\mathcal Z}\frac{h\tilde\nu^2N(z,\tilde\nu)\ c\ A_c}{4\pi d_l^2}\quad.
\ee
The observed frequency $\nu=\mathcal{D}\tilde\nu/(1+\mathcal Z)$ depends on the Doppler factor $\mathcal{D}$ and the redshift $\mathcal Z$. The flux from one cell is proportional to the photon density $N(z,\tilde\nu)$ and the surface of the cylindrical cell $A_c=2R\Delta z$, where $\Delta z$ is the cell size along $z$. The speed of light $c$ reflects the escape timescale given above.

In the time domain, light travel times have to be accounted for. The total flux from the source is then the sum over the fluxes $F_i$ from all $zn$ cells at the time, when they were causally connected to the reference point. This point is chosen to be the end point at $z=z_{max}$, i.e. closest to the observer.
\be
  \nu F_\nu^{tot}(\nu,t)=\sum_{i=1}^{zn} \nu F_i(z_i,\nu,t-t_i)
\ee
The time delay $t_i$ can be calculated from the additional distance along the line of sight
\be
  t_i=\frac{(z_{max}-z_i)\cos\theta_{los}}{c}\quad,
\ee
where $\theta_{los}$ is the angle between the $z$-axis and the line of sight. For time dependent simulations any time-interval in the observer's frame $\Delta t$ is boosted from the shock frame according to $\Delta \tilde t=\mathcal{D}\Delta t$.

\section{Results}
\label{sec:results}
In this section we will first present a classical SSC fit to the high energy data of the observational campaign by \citep[][]{2011ApJ...727..129A}. Based on the resulting fit we argue that the artificial spatial boundary condition should and can be replaced by a specific morphology of the radial confinement behind the shock.
\subsection{High energy SSC fit}
Similar to our previous work \textit{RS} the starting point is a classical SSC fit, meaning that the size of the emission region is a free parameter, shown as \textit{sim1} in Fig.~\ref{fig:low1Rlog}. In addition to the parameters in Table~\ref{tab:parameters} the parameters $r=2.5$, $D=\SI{E-15}{s^{-1}}$, an injection energy $\gamma_{inj}=50$ and an isotropization timescale $t_{iso}=\SI{1.1E4}{s}$ were used. These values were used for all simulations, except \textit{log2c}. The fit was done ``by eye'', as it is common for complex radiation codes.

An important parameter of any SSC fit is the synchrotron cooling break $\gamma_{cool}$, which determines the position of the the synchrotron peak flux. It is either explicitly set or consistently calculated from other parameters. Its position depends on the cooling time $t_{cool}$ and magnetic field strength $B$. The cooling time is closely connected to the numerical size of the emission region. In the case of our linear jet model $t_{cool}=z_{max}/(c\widetilde V_P)$. Using $\gamma_{cool}=(\beta t)^{-1}$, where $\beta$ is the synchrotron power, yields:
\be
  \gamma_{cool}=\frac{3m^3c^5}{2q^4}\frac{c\widetilde V_P}{z_{max}B^2}
\ee
If particles leaving the simulation region and their emission are neglected, then this dependency introduces an artificial boundary condition. This is the case for all SSC models that are either homogeneous or do not extend beyond a scale of $\sim\SI{E16}{cm}$.

Therefore we extend the emission region up to scales of ${\sim}\SI{1}{pc}$ (equivalent to $\SI{1.5}{mas}$) in the observer's frame and replace the numerical boundary condition by the shape of the function $R(z)$, as introduced in section~\ref{sec:radial_confinement}. The value of $z_{max}$ from Table~\ref{tab:parameters} corresponds to a length of $\mathcal{D} z_{max}\approx\SI{5.8}{pc}$ ($\SI{8.4}{mas}$).

\subsection{Morphology of the radial confinement}
\begin{figure}[ht]
  \centering
  \includegraphics[width=0.8\textwidth]{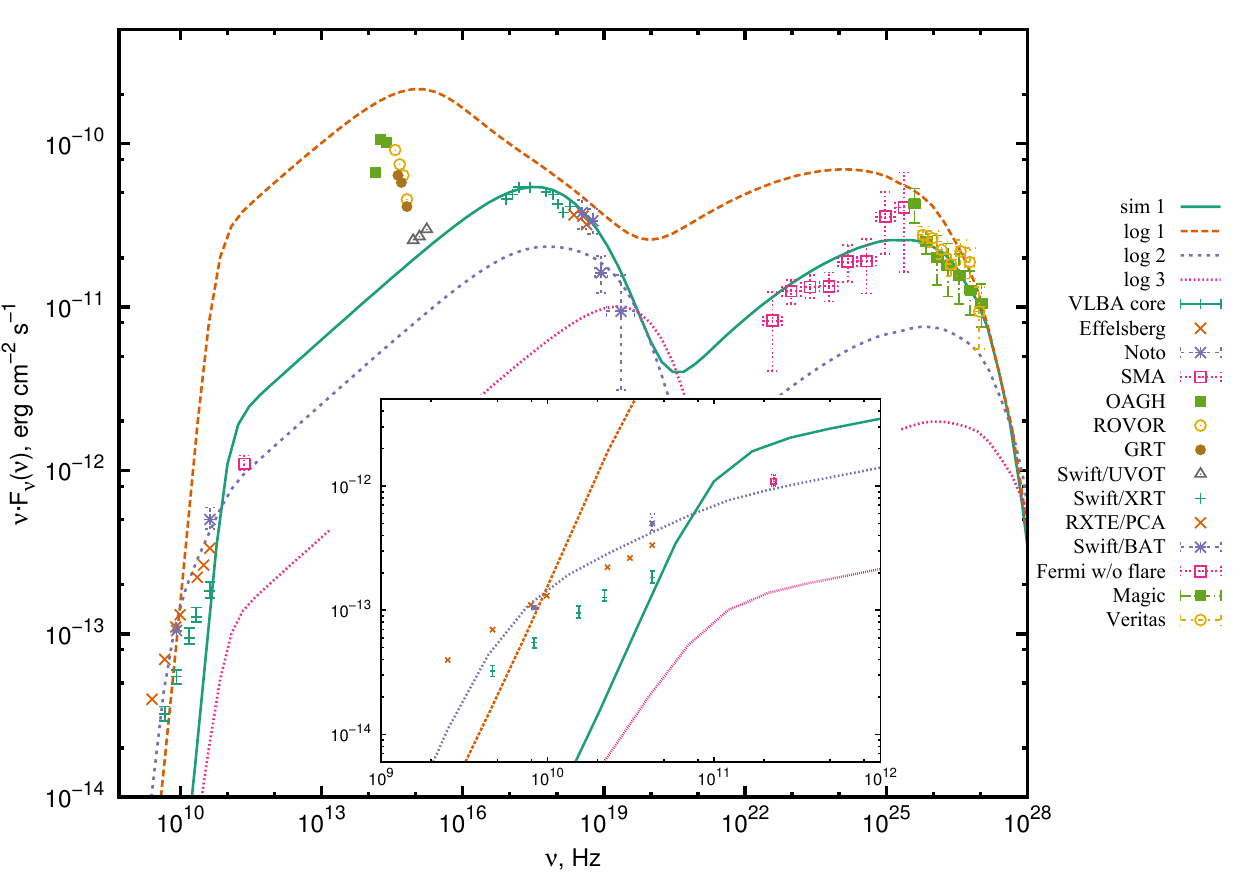}
  \caption{Shown are the overall SEDs for various simulations and a detailed view of the radio band. \textit{sim1} was obtained with a variable box size $z_{max}$, altering the resulting SED. Simulations \textit{log[1-3]} were obtained with a fixed, but large $z_{max}$ and show the effect of an increase in expansion speed. The peak in the optical is due to the host galaxy and can be fitted by a thermal component. Data taken from~\citep{2011ApJ...727..129A}.}
  \label{fig:low1Rlog}
\end{figure}

\begin{figure}[ht]
  \centering
  \includegraphics[width=0.8\textwidth]{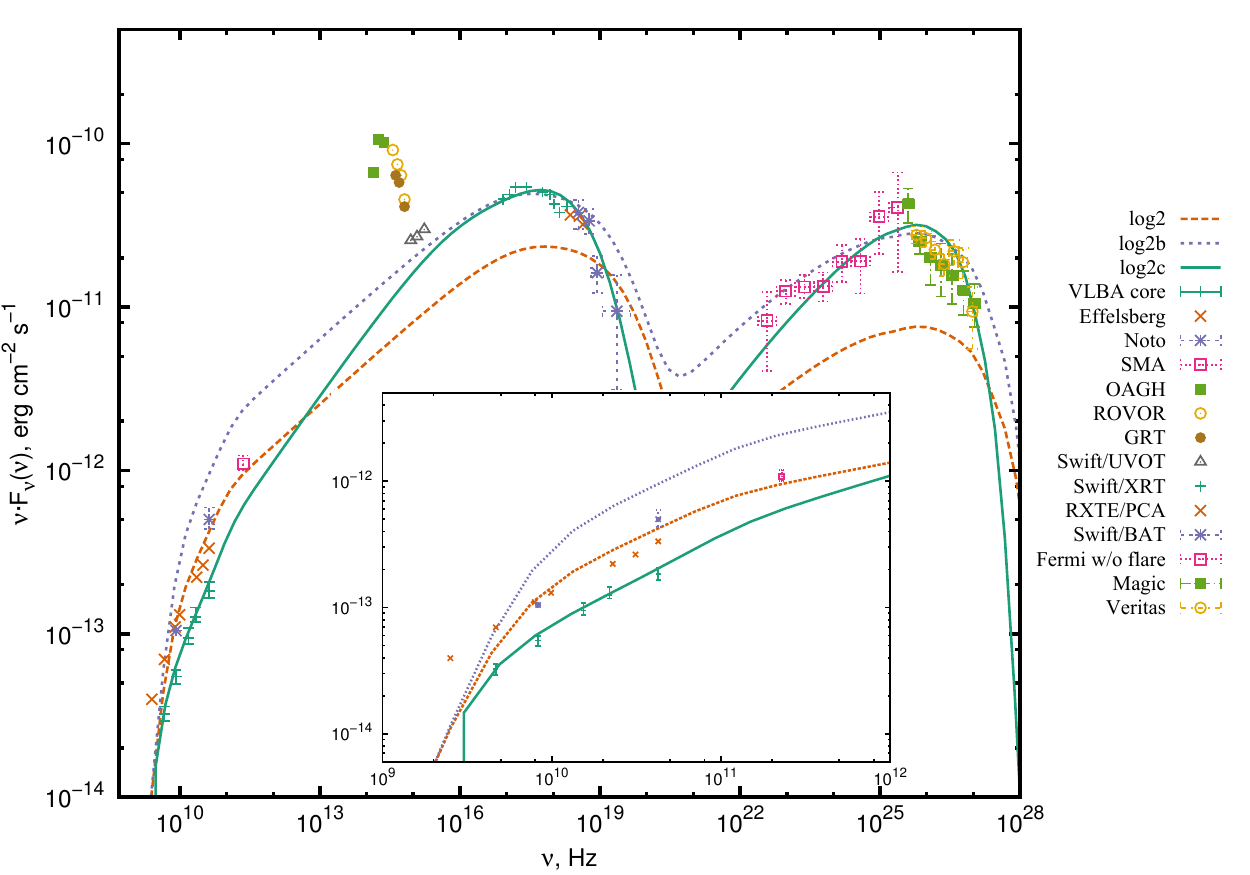}
  \caption{Flux adjusted SEDs after taking into account the effect of the radial expansion behind the shock. It is, however, obvious that a simultaneous fit of the X-ray, UV and radio data can not be achieved with an unbroken powerlaw, as it is expected from Fermi-I acceleration. Data taken from~\citep{2011ApJ...727..129A}.}
  \label{fig:newRlog}
\end{figure}

In order to illustrate the effect of the shape of $R(z)$, i.e. of the parameter $\alpha$, the simulations \textit{log1-3} were run with a range of values. Their SEDs are shown in Fig.~\ref{fig:low1Rlog}. The best representation of the radio spectral index is obtained with $\alpha=1$. The drop in flux in the high energy regime as well as an increase in the inverse Compton peak frequency $\nu_{IC}$ is due to the drop in density and magnetic field in the downstream vicinity of the shock. This can be compensated by a small change in the fit parameters, summarized in Table~\ref{tab:parameters}.
\begin{table}[h]
	\centering
	\caption{Parameters used for the curves in Figs.~\ref{fig:low1Rlog} and \ref{fig:newRlog}.}
	\begin{tabular}{crrrrr}
		\tableline\tableline
		\thead{Simulation} & \thead{$z_{max}$} & \thead{$B$} & \thead{$N_{inj}$} & \thead{$\mathcal{D}$} & \thead{$\alpha$}\\
		 & \tunit{cm} & \tunit{G} & \tunit{s^{-1}} & & \\
		\tableline
		\\[-5mm]
		sim1 & $\SI{8E15}{}$ & $\SI{0.023}{}$ & $\SI{4E43}{}$ & $45$ &  0 \\
		\textit{log1} & $\SI{4E17}{}$ & $\SI{0.023}{}$ & $\SI{4E43}{}$ & $45$ &  0.1 \\
		\textit{log2} & $\SI{4E17}{}$ & $\SI{0.023}{}$ & $\SI{4E43}{}$ & $45$ &  1 \\
		\textit{log3} & $\SI{4E17}{}$ & $\SI{0.023}{}$ & $\SI{4E43}{}$ & $45$ &  10 \\
		log2b & $\SI{4E17}{}$ & $\SI{0.025}{}$ & $\SI{6.8E43}{}$ & $48$ &  1\\
		log2c & $\SI{4E17}{}$ & $\SI{0.07}{}$ & $\SI{6.75E43}{}$ & $34.5$ &  1.5\\
		\tableline
	\end{tabular}
	\label{tab:parameters}
	\tablecomments{Simulations \textit{log[1-3]} are not fits in the strict sense, but merely a parameter scan for $\alpha$.}
\end{table}
The resulting SED \textit{log2b} is, however, overestimating the radio flux. Since the radio data is now constraining the overall fit, a solution with a much steeper slope \textit{log2c},  shown in Fig.~\ref{fig:newRlog}, is preferred. In addition to the ones in Table~\ref{tab:parameters} the following parameters were changed for the simulation \textit{log2c}: $r=3.5$, $t_{iso}=\SI{2E3}{s}$.

\begin{figure}[ht]
  \centering
  \includegraphics[width=0.7\textwidth]{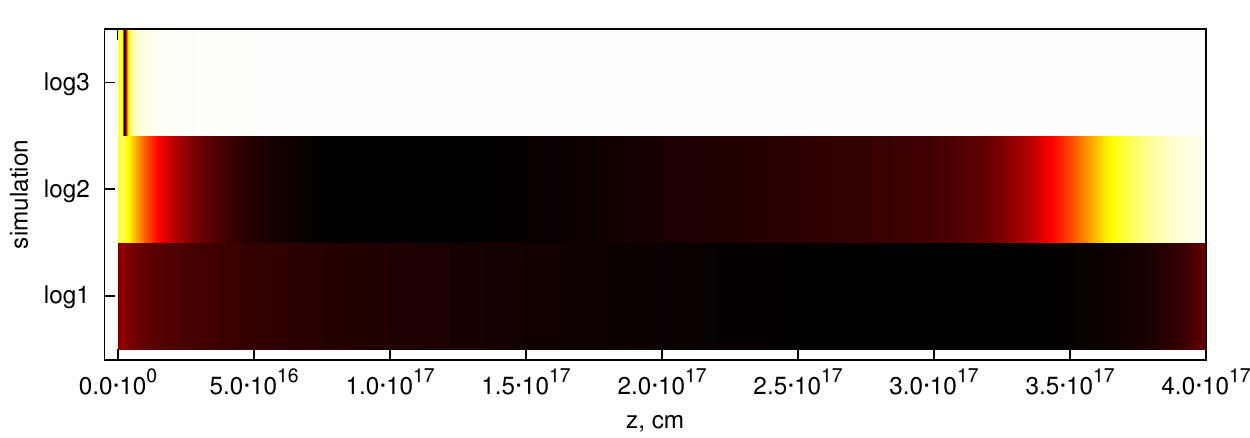}
  \caption{Flux depending on the distance from the shock at $\nu=\SI{4.4E09}{Hz}$. The same color-scale as in Fig.~\ref{fig:Rlog_morph} is used.}
  \label{fig:Rlog_morph_comb}
\end{figure}

\begin{figure}[ht]
  \centering
  \includegraphics[width=0.7\textwidth]{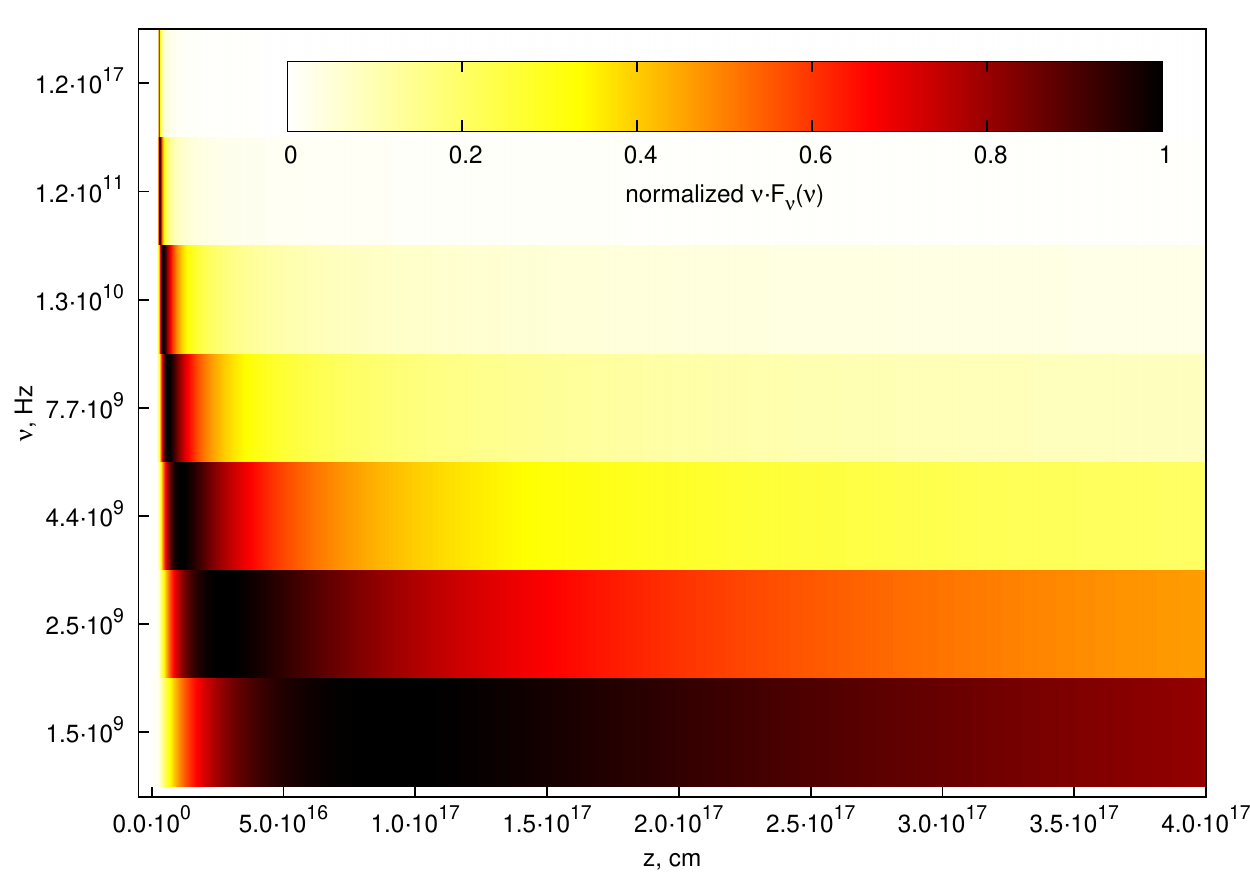}
  \caption{Fluxes produced by simulation \textit{log2c} for various frequencies with respect to the shock-distance in the shock-frame. The two different regimes (self-absorbed and not self-absorbed) can easily be distinguished.}
  \label{fig:Rlog_morph}
\end{figure}

The corresponding radio morphologies are shown in Figs.~\ref{fig:Rlog_morph_comb} and~\ref{fig:Rlog_morph}. The first plot shows the normalized flux at $\nu=\SI{4.4E09}{Hz}$ depending on the position within the emission region. The shock is positioned at $z_0=\SI{1.6E15}{cm}$. Hence, for large values, $z$ is approximately equal to the distance from the shock. A formation of a large scale radio structure is found for simulations \textit{log[1,2]}, but not for \textit{log3}.
In the second plot the change of the radio shape with respect to the frequency is shown for simulation \textit{log2c}.
\section{Discussion}
\label{sec:discussion}
The values for $\alpha$ and their effect on the SED represent three different regimes:

The first one is the almost homogeneous case, which will keep all physical parameters approximately constant and is reached in the limit $\alpha\rightarrow0$. In this case cooling will stay efficient even far downstream. This will lower the spectral break energy of the electron distribution and decrease the synchrotron peak frequency $\nu_{sync}$. As seen from simulation \textit{log1}, in this case no fit to the data can be obtained. This will hold for almost all Blazars, since the ratio $\Delta\nu_{peak}=\nu_{IC}/\nu_{sync}$ is one of the most significant feature of their SEDs. Furthermore, the radio part of the SED will still be fully self-absorbed.

The opposite case is represented by simulation \textit{log3}. For $\alpha\gg1$, the emission in all frequencies is dominated by the immediate vicinity of the shock. This corresponds to our results for large conical expansion presented in \textit{RS}. As can be seen in Fig.~\ref{fig:Rlog_morph_comb}, such strong expansion cannot connect the observed radio emission with the high energy emission.

The best representation of the radio data, as well as a strong boundary condition approximately preserving $\Delta\nu_{peak}$, is obtained with $\alpha=1$. The resulting flux morphology is presented in Fig.~\ref{fig:Rlog_morph}. The transition to the self-absorbed regime at around $\SI{100}{GHz}$ is well reflected in the morphology. While the emission for optically thin frequencies is dominated by the vicinity of the shock, the peak flux for self-absorbed frequencies is much further downstream. The core shift, usually observed in AGN jets~\citep{1998A&A...330...79L}, as well as the increasing size of the radio blob with decreasing frequency is reproduced in our model. These flux distributions could in principle be matched against VLBI radio maps.

Furthermore it is important to note, that the radio data, when taken into account, strongly constrain the overall fit. As can be seen from the SED of \textit{log2c} in Fig.~\ref{fig:newRlog} the spectral index of the rising synchrotron flank is changed, which in turn influences the spectral index in the Fermi band via the inverse Compton process.

Here the overall fit suffers from the problem that the slopes below $\nu_{sync}$ and $\nu_{IC}$ do not have the same spectral index, as expected from the SSC paradigm. This problem was already discussed by e.g. \citet{2011ApJ...740...64L,2011ApJ...743L..19L}. The connection to the radio spectrum can now be used to distinguish the different fit solutions. In our model a flat synchrotron spectrum can not be arranged with the observed radio fluxes, favoring a fit that still represents the \textit{Fermi}-data, but omitting the data of \textit{UVOT} and \textit{SMA}. A complete fit could be achieved by two additional breaks, where the first one, around $\SI{E11}{Hz}$, would be towards a higher spectral index, or by additional components. A possible explanation for the \textit{UVOT} data could, for a low state of the jet component, be the emission from the broad line region (BLR), as discussed by~\citet{2012MNRAS.420.2899G}. The discrepancy between \textit{SMA} and \textit{VLBA} could be due to the fact that the latter has a resolution roughly four orders of magnitude higher. Hence the emitting region of the flux detected by \textit{SMA} is orders of magnitudes larger than our simulation box.

\section{Conclusion}
The consistent connection of high energy and radio emissions from AGN is an important step towards a deeper understanding of these sources. Multi-frequency observation campaigns yield a rich database of radio data and correlations with various bands. So far, no models exist to test this data against.

The model presented in this work is able to track the particle distribution, responsible for the VHE emission far downstream up to scales of ${\sim}\SI{1}{pc}$. The full time-dependence of the model is achieved by the explicit implementation of Fermi-I acceleration. The (steady state) conclusions in this work are however independent from the actual acceleration mechanism. The lightcurves resulting from various flare scenarios will be summarized in a subsequent publication.

We find the formation of a radio structure downstream of the acceleration site to depend heavily on the morphology of the confinement of the non-thermal energy distribution. The conservation of sufficiently small variability timescales requires a strong expansion behind the shock on length scales typical for homogeneous SSC models. Subsequently, the development of a radio structure requires a slowdown of the expansion. As a qualitative ansatz that fulfills these requirements and produces a good fit to the VHE as well as the radio part of the SED a log function was used.

The neglect of photon propagation between cells can be justified by the small gradient of the photon densities in the pure SSC case. A photon diffusion term following~\citep{2011ApJ...727...21J} was implemented and found to only have marginal influence on the produced SED. Therefore it was neglected for the sake of computational efficiency. In the case of extreme time variability, second order inverse Compton or external Compton scattering this simplification should be re-evaluated, as indicated by previous work~\citep[e.g.][]{2008ApJ...689...68G}.

The current limitations of our model are the assumed homogeneity in radial direction as well as the isotropic approximation for all radiation processes. These should not alter the conclusions in this work, but prevent the usage of additional data, most prominently the electric vector polarization angle (EVPA) and its correlation to lightcurves. A statistical approach to this problem was presented by~\citet{2014ApJ...780...87M}. The restriction to a one-dimensional representation will only allow a qualitative connection between the flux morphology predicted by our model and VLBI images of the associated radio structures. More importantly, the one-dimensional particle distribution restricts us to non-relativistic shocks. Also, considering the wealth of data, the current model should be seen as a proof of concept rather than a fitting algorithm for daily use.

Despite these limitations it was shown that the consideration of the radio emission can distinguish between different sets of parameters, thus increasing our understanding of these sources. Furthermore a correlation between VHE flares and a disturbance of the radio morphology could be used to identify the acceleration site of the radiating particles.

\acknowledgments
We would like to thank the referee for his helpful comments.

F.S. acknowledges support from NRF through the MWL program. This work is based upon research supported by the National Research Foundation and Department of Science and Technology. Any opinion, findings, and conclusions or recommendations expressed in this material are those of the authors and therefore the NRF and DST do not accept any liability in regard thereto.

S.R. wants to thank the Leibniz-Rechenzentrum (LRZ) for the provision of numerical resources.

\bibliography{apj-jour,ref}

\listofchanges

\end{document}